\documentclass[prb,twocolumn,nofootinbib, showpacs,preprintnumbers,amsmath,amssymb]{revtex4}
\usepackage{graphicx}
\usepackage{dcolumn}

\begin{document}

\title{Effects of cluster diffusion on the   island density and size distribution in submonolayer island growth }

\author{Y.A. Kryukov}
\author{Jacques G. Amar}

\affiliation{Department of Physics and Astronomy \\ University of Toledo,  Toledo, Ohio 43606, USA}

\begin{abstract}

The effects of    cluster diffusion on the   submonolayer island density $N$   and island-size distribution  $N_s(\theta)$ (where $N_s$ is the density of islands of size $s$ at coverage $\theta$)
are studied for the case of irreversible growth of compact islands   on a 2D substrate.
 In our model, we assume instantaneous coalescence of circular islands, while  the mobility $D_s$ of an island of size $s$
 (where $s$ is the number of particles in an island) satisfies $D_s  \sim s^{-\mu}$.  Results are presented for    $\mu = 1/2$ (corresponding to Brownian motion),
 $\mu =1$ (corresponding to correlated evaporation-condensation),  and $\mu = 3/2$ (corresponding to cluster diffusion via edge-diffusion),  as well as for higher values   including $\mu = 2,3,$ and $6$.  We also compare our results with those obtained in the limit of no cluster mobility ($\mu = \infty$).
In general, we find that the exponents $\chi$ and $\chi'$
describing the flux-dependence of the island and monomer densities respectively,
vary continuously as a function of  $\mu$.
Similarly, the exponent $\omega$ describing the flux-dependence of the coverage $\theta_m$ corresponding to the peak island-density also depends continuously on $\mu$, although the exponent $\omega'$ describing the flux-dependence of the coverage corresponding to the peak monomer density does not.
In  agreement with theoretical predictions that  for point-islands with $\mu < 1$   power-law behavior
of the island-size distribution (ISD)
is expected,
for $\mu = 1/2 $ we find   $N_s \sim s^{-\tau}$
up to a cross-over island-size $S_c$.
However, the value of the exponent $\tau$ obtained in our simulations ($\tau \simeq 4/3$) is  higher than the point-island  prediction $\tau = (3 - \mu)/2$. Similarly, the measured value  of the exponent  $\zeta$ corresponding to the dependence of
$S_c$ on the average island-size $S$ (e.g. $S_c \sim S^\zeta$) is also significantly higher than the point-island prediction  $\zeta = 2/(\mu+1)$.
For  $\mu < 1$,
a   generalized   scaling form for the ISD,  $N_s(\theta)  = \theta/S^{1+\tau \zeta}   f(s/S^\zeta)$,   is also proposed, and using this form excellent scaling of the entire distribution is found for $\mu  = 1/2$.
However, for finite $\mu \ge 1$ we find that,
due to the competition between two different size-scales,
neither the generalized scaling form nor the standard scaling form $N_s(\theta)  = \theta/S^{2}   f(s/S)$ lead to
scaling of the entire ISD for finite values of the ratio $R = D_1/F$ of the monomer diffusion rate to deposition flux.
Instead, we find that  the scaled ISD becomes more sharply peaked with increasing $R$ and coverage.
This is in contrast to models of epitaxial growth with limited cluster mobility for which
 good scaling occurs  over a wide range of coverages.

\end{abstract}


\maketitle
\section{Introduction}

Recently, there has been a lot of interest in understanding the scaling behavior in submonolayer island nucleation and  growth.\cite{Walton,Venables73, Venables84,Kunkel,Zhang,Evansreview}
One reason for this is that the submonolayer growth  regime plays an important role in determining the later stages of thin-film growth.\cite{Walton,Venables73, Venables84,Kunkel,Zhang}
Of particular interest is the dependence of the  total
island-density  $N$ and island-size distribution $N_s(\theta)$ (where $N_s$ is the   density of islands of
size $s$ at coverage $\theta$ and $s$ is  the number of monomers in an island)
on   deposition parameters such as the
 deposition flux $F$
  and growth temperature $T$.

One concept that has proven especially useful in studies of submonolayer epitaxial growth
is that of a critical island size,\cite{Venables73} corresponding to one less than the size of the smallest  ``stable" cluster.
For example, if we  assume that only monomers can diffuse,  then in the case of  submonolayer growth of 2D islands on a solid 2D substrate,  standard nucleation theory\cite{Venables73, Venables84} predicts that the peak island density $N_{pk}$ and the  monomer density $N_1$ at fixed coverage  satisfy,
\begin{equation}
N_{pk} \sim (D_{1,h}/F)^{-\chi_i} ~~~~~N_{1} \sim (D_{1,h}/F)^{-\chi'_i}
\label{scaling}
\end{equation}
 where $D_{1,h}$ is the monomer hopping rate,
 $i$ is the critical island size,  $\chi_i =  \frac {i}  {i+2}$ and $\chi'_i  = 1 - \chi_i$.
We note that in the case of irreversible island growth ($i = 1$) this implies that  $\chi_1 = 1/3$ and $\chi'_1 = 2/3$.
In addition,   it has been shown  that  in the absence of cluster-diffusion and in
the pre-coalescence regime    the island-size distribution (ISD) satisfies the scaling form,
\cite{BE94, Amar94}
\begin{equation}
N_s(\theta)= \frac{\theta}{S^2}f_i\left(\frac{s}{S}\right),
\label{isdscal}
\end{equation}
where $S$ is the average island size, and the scaling function $f_i(u)$ depends
on  the critical island size.\cite{iprl}

However,  in some  cases (such as in epitaxial growth on metal(111) surfaces) it is also possible for significant {\it small} cluster diffusion to occur.\cite{crc, gnandip,  Rahman1}
 In addition,
several   mechanisms for the diffusion of {\it large} clusters on solid surfaces have also been proposed.
\cite{Wen,Siclen, Khare,Shollprl,Sholl, Soler, Pai}
In each case, scaling arguments predict that the cluster diffusion coefficient $D_s$ decays as a power-law with island-size $s$ (where $s$ is the number of particles in a cluster), i.e. $D_s \sim s^{-\mu}$.
In particular, three different limiting cases have been  considered\cite{Wen,Siclen, Khare,Shollprl,Sholl, Soler, Pai}  - cluster diffusion due to    uncorrelated evaporation-condensation  ($\mu = 1/2$),   cluster diffusion due to    correlated evaporation/condensation ($\mu = 1$), and  cluster diffusion due to  periphery diffusion  ($\mu = 3/2$).  We note that the case $\mu = 1/2$ also corresponds to the Brownian (Stokes-Einstein) diffusion of compact 2D clusters in two-dimensions.

 In order to   understand   the effects of island diffusion on the submonolayer scaling behavior a number of simulations have previously been carried out.  For example, Jensen et al\cite{Jensen} have studied the effects of  island-diffusion with $\mu = 1$ on the   percolation coverage for the case of    irreversible growth without relaxation, corresponding to islands with fractal dimension $d_f \simeq 1.5$.   More recently,  Mulheran and Robbie\cite{Mulheran} have used a similar   model   to study the dependence of the exponent $\chi$
 on the cluster-diffusion exponent $\mu$ for values of $\mu$  ranging from $ 0$ to $9$.   They found that for small values of $\mu$ the value of the exponent  ($\chi \simeq 0.45$)  is significantly larger than the value   ($\chi \simeq 1/3$) expected in the absence of cluster diffusion, although it  decreases with increasing $\mu$.
  However,  the scaling of the ISD was not studied.\cite{Kuipers-sim}

Motivated in part by these simulations,  Krapivsky et al\cite{Krapivsky1,Krapivsky2} have   carried out an analysis of the scaling behavior for the case of point-islands,
based on the corresponding mean-field Smoluchowski equations.\cite{Smol}
Their analysis suggests that  due to the large amount of diffusion and coalescence in this case, for $\mu < 1$
the total island density saturates (corresponding to ``steady-state" behavior) while
the ISD  exhibits  power-law  behavior  of the form, $N_s \sim   s^{-\tau}$, where $\tau = (3-\mu)/2$ and the prefactor   does not depend on coverage.\footnote{The   expression $\tau = (3 - \mu)/2$
has also been derived by Cueille and Sire\cite{Cueille} and Camacho.\cite{Camacho}}
This power-law dependence for the ISD is predicted to  hold
up to a critical island-size $S_c$, where $S_c \sim S^\zeta$ and $\zeta = 2/(\mu + 1)$.
In contrast, for $\mu \ge 1$  continuous island evolution is  predicted, e.g. the total island density does not saturate, and as a result  no simple power-law behavior is predicted for the ISD.
Their analysis also indicates that  for all values of $\mu$, one has $\chi_1 = \chi'_1 = 1/2$ with logarithmic corrections.
However, it should be noted that the point-island approximation is typically only valid at extremely low coverages.

Here we present the results of kinetic Monte Carlo simulations of
   irreversible island growth with cluster diffusion  for the case of compact islands with fractal dimension $d_f = 2$.  Among the primary motivations for this work are  recent experiments\cite{Bigioni} on the growth of (compact) colloidal nanoparticle islands
at a liquid-air interface in which
significant cluster diffusion has been observed.
Accordingly, in contrast to  much of the previous work\cite{Jensen, Mulheran, Kuipers} our model is an off-lattice model.
However, our main goal here is not to explain these experiments but rather to obtain results
which may be used as a reference for future work.
As already noted, if cluster diffusion is due to 2D Brownian motion (as might be expected at a fluid-interface) then the value of the exponent $\mu$
($\mu = 1/2$) is the same as that expected for uncorrelated evaporation-condensation.  However,   we also present results for $\mu = 1$ (corresponding to cluster-diffusion due to correlated evaporation-condensation), $\mu = 3/2$ (corresponding to cluster-diffusion due to periphery diffusion)  as well as for higher values of $\mu$ ($\mu = 2,3,6$ and $\infty$).

This paper is organized as follows.  In Sec.~II, we     describe our model  in detail along with the parameters used in our simulations, while in Sec.~III we discuss    the methods we have used to enhance the simulation efficiency.  In Sec.~IV we   derive a generalized scaling form for the ISD which is appropriate for the case of a power-law ISD with $\tau > 1$,  corresponding to $\mu < 1$.  We then present our results for the scaling of the island-size distribution and island and monomer densities  as a function of $D_{1,h}/F$, coverage, and $\mu$  in Sec.~V.  Finally, in Sec.~VI we   discuss our results.

\section{Model and Simulations }

For simplicity we have studied a model of irreversible aggregation in which all islands are assumed to be circular and    rapid island   relaxation (perhaps due to  periphery diffusion) is assumed.  In particular, in our model  each island or cluster  of size $s$ (where $s$ is the number of monomers in a cluster) is represented by a circle with area $A_s = \pi d_s^2/4 $ and diameter $d_s =  d_1 s^{1/2}$, where $d_1$ is the monomer diameter. In addition, each cluster of size $s$ may diffuse with  diffusion rate $D_s = D_1 s^{-\mu}$ where $D_1 =  D_{1,h} \delta^2/4$ is the monomer diffusion rate, $D_{1,h}$ is the monomer ``hopping rate", and $\delta$ is the hopping length.
Similarly, we may write $D_s = D_{s,h} ~\delta^2/4$ where $D_{s,h}$ is the hopping rate for a cluster of size $s$.

In order to take into account  deposition,
monomers are also randomly deposited onto the   substrate with rate $F/d_1^2$ per unit time per unit area.  Since instantaneous coalesce and relaxation is assumed, whenever  two clusters  touch  or overlap,  a new island is formed whose area is equal to the sum of the areas of the original clusters, and whose center corresponds to the center-of-mass of both islands.  We note that in some cases  a coalescence event may lead to overlap of the resulting cluster with additional clusters.  In this case, coalescence is allowed to proceed until   there are no more overlaps.
In addition,  if a monomer lands on an existing cluster, then that monomer is automatically `absorbed' by the cluster.

Thus, at each step of our simulation either a monomer is deposited (followed by a check for overlap with any clusters) or a cluster is selected for diffusion.   If a cluster is selected for diffusion, then the center of the cluster is displaced by a distance $\delta$ in a randomly selected direction.  For computational efficiency, and also because it is the smallest length-scale in the problem, in most  of the results presented here we have assumed $\delta = d_1$.  However, we have also carried out some simulations  with  smaller values   ($\delta=0.5 ~d_1$ and $\delta=0.25 ~d_1$) in order to approach the continuum limit.
As discussed in more detail in Sec.~VI, our results indicate that the dependence of the island   and monomer densities on the hopping distance $\delta$ is relatively weak.

We note that besides the exponent $\mu$ describing the dependence of the cluster diffusion rate on cluster-size, the other key parameter in our simulations is the ratio $R_h$ of the monomer hopping rate to the monomer deposition rate (scaled by the ratio of the hopping length to the monomer diameter)  e.g.,
\begin{equation}
R_h = \frac {D_{1,h}}{F} \left (\frac  {\delta}{d_1}\right)^2
\end{equation}
We note that this definition implies that the dimensionless ratio $R = D_1/F d_1^2$ of the monomer diffusion coefficient $D_1$  to the deposition flux satisfies,
\begin{equation}
R = R_h/4
\end{equation}

Our simulations were carried out assuming a 2D square substrate of size $L$ (in  units of the monomer diameter $d_1$) and periodic boundary conditions.  In order to avoid finite-size effects, the value of $L$ used ($L = 4096$) was relatively large, while our results were averaged over  $100$ runs in order to obtain good statistics. In order to determine the asymptotic dependence of the island density on coverage and $R_h$ our simulations were carried out using values of $R'_h=4 R_h/\pi$ ranging from $10^7 - 10^{9}$ up to a maximum coverage of $0.3$ monolayers (ML).  In order to study the dependence on $\mu$,  simulations were carried out  for  $\mu = 1/2$ (corresponding to Brownian diffusion or uncorrelated evaporation-condensation), $\mu = 1$ (corresponding to correlated evaporation-condensation), and $\mu  = 3/2$ (corresponding to periphery diffusion) as well as for higher values ($\mu = 2,3,$ and $6$) as well as the case $\mu = \infty$ corresponding to only monomer diffusion.

 In order to obtain a quantitative understanding of the submonolayer growth behavior, we have measured a variety of quantities  including the monomer density $N_1 = (\pi/4)~ n_1/L^2$ (where $n_1$ is the number of monomers in the system) as a function of coverage $\theta$, and the total island density $N  = (\pi/4)~n/L^2$ (where $n$ is the total number of islands including monomers in the system).   In addition, we have also measured the island-size distribution $N_s(\theta)$ where $N_s = (\pi/4)~n_s/L^2$ corresponds to the density of islands of size $s$.
We note that the factors of $\pi/4$ in the definitions above take  into account the fact that the area of a monomer is $(\pi/4)~d_1^2$, and as a result the densities defined above all correspond to area fractions.
 Similarly,  the coverage     $\theta  =  \sum_{s\ge 1} s N_s$    corresponds to the fraction of the total area covered by islands (including monomers).

\section{Simulation Methods}

While a simple Monte Carlo approach can be used\cite{Meakin} to simulate the processes of monomer deposition and cluster diffusion such a method can be very inefficient for large values of $R_h$ and small values of $\mu$, since the large range of island-sizes and diffusion rates can lead to a low acceptance ratio.  Accordingly, here we use  a kinetic Monte Carlo approach.   In particular, if we set the deposition rate $F$ per unit area $d_1^2$ equal to $1$, then the total deposition rate in the system is $L^2$ while the hopping rate for a cluster of size $s$ is given by $R_{s,h} = R_h s^{-\mu}$. As a result, the total diffusion rate for all clusters is given by $R_T = \sum_{s=1}^\infty n_s R_{s,h}$ (where $n_s$ is the number of clusters of size $s$) while the total rate of deposition onto the substrate is $L^2$.  The probability $P_{dep}$ of depositing a monomer is then given by,
\begin{equation}
P_{dep} = \frac {L^2} {R_T + L^2}
\end{equation}
while the probability of cluster diffusion is $P_{diff} = 1 - P_{dep}$.  If cluster  diffusion is selected, then a binary tree\cite{Blue} (whose bottom leaves correspond to the  total hopping rate  $n_s R_{s,h}$ for each size $s$)  may be used to efficiently select with the correct probability which cluster will move as well as to efficiently update $R_T$.
However, for large $R_h$ and small $\mu$ the maximum cluster-size  can be  larger than $10^4$ and as a result the computational overhead associated with the binary tree can still be significant.

Accordingly, we have implemented a variation\cite{Schulze} of the binary tree approach in which a range of cluster-sizes are clustered together into a single `leaf' or bin.
In particular, to minimize the size of the binary tree, starting with island-size $s > 3$  we have used  variable bin-sizes such that each bin  contains several different cluster  sizes ranging from a starting value $i$ to a value approximately equal to $1.2 i$.
Using this scheme allows us to use a binary tree with a maximum of $64$ leaves and a rejection probability of only $10\%$.  To further decrease the computational overhead, our binary tree  grows dynamically from $4$ leaves to as many as needed.

By properly selecting the rates in the binary tree and the corresponding acceptance probabilities,  one can ensure that each diffusion event is selected with the proper rate.  In particular, if  we define the rate of bin $i$ as ${\cal R}_{b_i}=n_{b_i} {\cal R}_{max,b_i}$, where ${\cal R}_{max,b_i}$ is the maximum cluster-diffusion rate in   bin $b_i$ (corresponding to the smallest cluster-size in the bin) and $n_{b_i}$ is the number of islands in the bin, then the sum over all leaves may be written,
\begin{equation}
{\cal R}_b^{T}=\sum_{i=1}{\cal R}_{b_i},
\label{Ptot_kmcb}
\end{equation}
The probability of attempting a diffusion event is then given by,
\begin{equation}
P_{diff} = \frac {{\cal R}_b^{T}}{{\cal R}_b^{T} +L^2}
\label{Pdiffbin}
\end{equation}
while the probability of selecting bin $i$ is given by $P_i = {\cal R}_{b_i}/{\cal R}_b^T$.
Once a bin is selected using the binary tree, a specific cluster is then selected randomly from the list of all the clusters in that bin. This implies that  a cluster of size $s$ will be selected with probability $P_s = n_s/n_{b_i}$.
Thus, by assuming an acceptance probability for the selected cluster-diffusion event given by
\begin{equation}
P_{acc}=\frac{R_s}{{\cal R}_{b_i}},
\label{pacceptance}
\end{equation}
each diffusion event will be selected with the proper rate.

Since our simulations are carried out off-lattice, one of the most time-consuming processes  is the search for overlaps every time a cluster is moved. While the simplest way to carry out such a search is to check for overlaps with all other islands in the system,  the search time scales as $L^2$, and as a result it becomes very time-consuming for large systems.  Accordingly,  we have used a neighbor look-up table\cite{liquids} which contains  a list of all other islands within a buffer-distance of each island. The search for overlaps is then carried out only among the neighbors on this list rather than over all the islands in the system. The neighbor list is updated whenever the total displacement   of any island since the last update is larger than half the buffer-distance.

To speed-up the updates of the neighbor table, we have also used a ``grid'' method\cite{liquids} in which our system is divided into    an $n_g$ by $n_g$ grid of boxes of size $l_g = L/n_g$ and each cluster can be rapidly assigned to a given box. Using this method the search for neighbors only includes clusters within an island's box as well as the  $8$ adjacent boxes.  As a result, the table update time is reduced to $9 L^2/n_g^2$ instead of $L^2$. To further optimize the speed of our simulations, the grid size is varied as the average island-size increases.

\section{Generalized scaling form for the Island-size Distribution}

As discussed in Sec.~I, in both simulations and experiments on submonolayer epitaxial growth, the island-size distribution (ISD) is typically assumed to satisfy the scaling form given in Eq.~\ref{isdscal}.
However, this scaling form has been derived\cite{BE94, Amar94, iprl} on the assumption that there is only one characteristic size-scale $S$ corresponding to the average island-size, and that the ISD does not diverge for small $s/S$.  However, in our simulations of monomer deposition and cluster diffusion and aggregation with   $\mu < 1$, we find that the ISD exhibits a well-defined power-law behavior for small $s/S$.  In addition,   the existence of a  shoulder in the ISD for large $s = S_c$ implies the existence of  a second characteristic length-scale which scales as $S_c \sim S^\zeta$. 
We note that this corresponds to an island size-scale such that steady-state behavior breaks down, due to the existence of   mass-conservation and a finite diffusion length. 

In general one would expect this to lead to a more complicated two-variable scaling of the form $N_s (\theta) = A ~g(s/S, s/S_c)$.  However, if the power-law behavior for small $s/S$ is well-defined (and $\tau > 1$) then it is  possible to derive a   generalized scaling form involving only one variable.  In particular, we assume that a scaling form for the island-size distribution may be written,
\begin{equation}
N_s(\theta)  = A(S,\theta)  f(s/S^\zeta)
\end{equation}
In order to determine $A(S,\theta)$  note that  $N=\sum_{s\ge1} N_s=\theta/S=A(S,\theta) \sum_{s\ge1} f(s/S)\Delta s$. Converting to an integral this may  be rewritten as $\theta/S=A(S,\theta) ~S^\zeta~\int_{1/S^\zeta}^\infty  f(u)~ u ~du$ where $u=s/S^\zeta$. If we now assume that $f(u) \sim u^{-\tau}$ for small $u$ and $\tau > 1$,  then the small-$u$ part of the integral dominates and we obtain, $A=\theta/S^{1+\tau\zeta}$.  This leads to the generalized scaling form,
\begin{equation}
N_s(\theta)=\frac{\theta}{S^{1+\tau\zeta}} f\left(\frac s{S^{\zeta}}\right),
\label{NS2Q}
\end{equation}
 We   note that  a similar scaling form  (corresponding to the special case $\zeta = 1$)    has previously been derived   in Ref.~\onlinecite{fm} for the case of the deposition of spherical droplets  with dimension $D > d$ on a $d$-dimensional substrate.
 We also note that for $\zeta = 1$ and $\tau = 1$ (corresponding to the critical value of $\tau$) the standard scaling form Eq.~\ref{isdscal} is obtained.

\section{Results}

\subsection {Stokes-Einstein diffusion ($\mu = 1/2$)}

\begin{figure}[t]
\includegraphics [width=7.5cm]{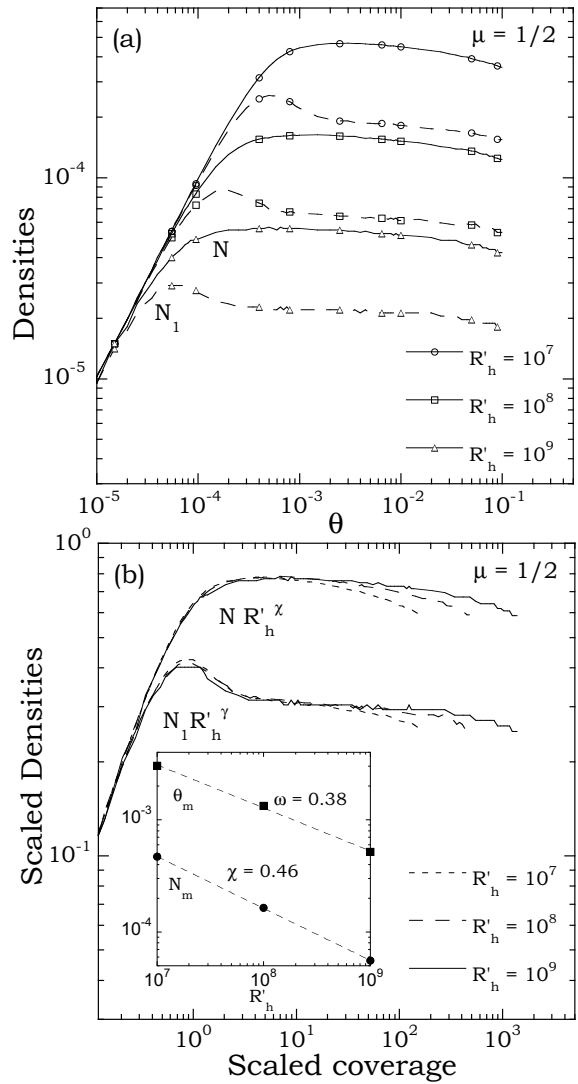}
\caption{(a) Island and monomer densities $N$ and $N_1$  as a function of coverage $\theta$ for ${R'_h}=10^7 - 10^9$ and $\mu = 1/2$. (b) Scaled densities $N {R'_h}^{\chi}$ and $N_1{R'_h}^{\gamma}$ as a function of scaled coverage  ($\theta {R'_h}^{\chi}$ and $\theta {R'_h}^{\gamma}$, respectively).  Inset shows dependence of peak island density $N_m$ and   coverage $\theta_m$ on $R'_h$.}
\label{Fig:dens05}\end{figure}

We first consider the case $\mu  = 1/2$ corresponding to Stokes-Einstein diffusion.
Fig.~\ref{Fig:dens05}(a) shows our results for the total cluster density $N$ (including monomers) as well as for the monomer density $N_1$ as a function of coverage for three different values of $R'_h$ ranging from $10^7$ to $10^9$.  In good agreement with the theoretical prediction in Refs.~\onlinecite{Krapivsky1} and \onlinecite{Krapivsky2} of   ``steady-state" behavior for $\mu < 1$,  we find that both the monomer density $N_1$ and total island density $N$ reach an approximately constant value beyond a critical coverage $\theta_m$.  We note that this coverage decreases with increasing $R_h$, while the peak island and monomer densities also decrease  with increasing $R_h$.

The inset in Fig.~\ref{Fig:dens05}(b) shows our results for the exponents  $\chi$ ($\chi \simeq 0.46$) and $\omega$ ($\omega \simeq 0.38$) corresponding to the dependence of the peak island density $N_m$ and   coverage $\theta_m$   on $R'_h$.
In qualitative agreement with the   results of  Mulheran et al\cite{Mulheran} for   fractal islands,  the  value of $\chi$ obtained in our simulations is  slightly lower but close to $1/2$.   This is also consistent with the prediction\cite{Krapivsky1,Krapivsky2} that for point-islands $\chi$ should be equal to $1/2$ with logarithmic corrections.
Fig.~\ref{Fig:dens05}(b) shows the corresponding scaled island  density $N  {R'_h}^\chi$  as a function of the scaled coverage $\theta {R'_h}^\chi$.  As can be seen there is  good scaling  up to and even somewhat beyond the value ($\theta {R'_h}^\chi \simeq 1$) corresponding to the peak in the island-density.   In contrast, replacing the scaled coverage by  $\theta {R'_h}^\omega$ as  in Ref.~\onlinecite{Mulheran},  leads  to good scaling at $\theta = \theta_m$, but the scaling is significantly worse for $\theta \neq \theta_m$.
Also shown is the scaled monomer density $N  {R'_h}^\gamma$ (where the peak monomer density scales as $N_{1,pk} \sim {R'_h}^{-\gamma}$ and the coverage corresponding to the peak monomer density scales as $\theta_{1,m} \sim {R'_h}^{-\omega'}$ and $\gamma \simeq \omega' \simeq 1/2$) as a function of the scaled coverage $\theta {R'_h}^\gamma$.  As for the case of the island density, there is good scaling up to and even beyond the scaled coverage corresponding to the peak of the monomer density.
We note that in contrast to the exponents $\chi$ and $\omega$,  the exponent $\gamma$ does not appear to depend on $\mu$.  In particular, we find that for all the values of $\mu$ that we have studied,  the value of $\gamma$ ($\gamma \simeq 0.45 - 0.47$)   is close to the value ($\gamma = 1/2$) expected in the absence of cluster-diffusion.

We now consider the scaled island-size distribution (ISD).  In Refs.~\onlinecite{Krapivsky1} and \onlinecite{Krapivsky2}  ``steady-state" power-law behavior of the form,
\begin{equation}
N_s(\theta) \sim s^{-\tau} {R'_h}^{-\chi}
\label{steadystate}
\end{equation}
where  $\tau = (3-\mu)/2$  was predicted for $\mu < 1$   for island-sizes $s << S_c$ where $S_c$ corresponds to the shoulder in the ISD for large $s$.    Similarly, the exponent $\zeta$ characterizing the scaling of $S_c$ as a function of $S$ (e.g. $S_c \sim S^\zeta$) was predicted to satisfy  the expression $\zeta =  2/(\mu+1)$.
We note that for $\mu= 1/2$ these expressions imply that $\tau = 5/4$ and $\zeta = 4/3$.
Since $N \sim {R'_h}^{-\chi}$ and $S = \theta/N$, one has  $\theta/S \sim {R'_h}^{-\chi}$.   Accordingly, Eq.~\ref{steadystate} may be rewritten as,
\begin{equation}
N_s(\theta) \sim s^{-\tau}~ \theta/S
\label{steadystate2}
\end{equation}

\begin{figure}[t]
\includegraphics [width=7.5cm]{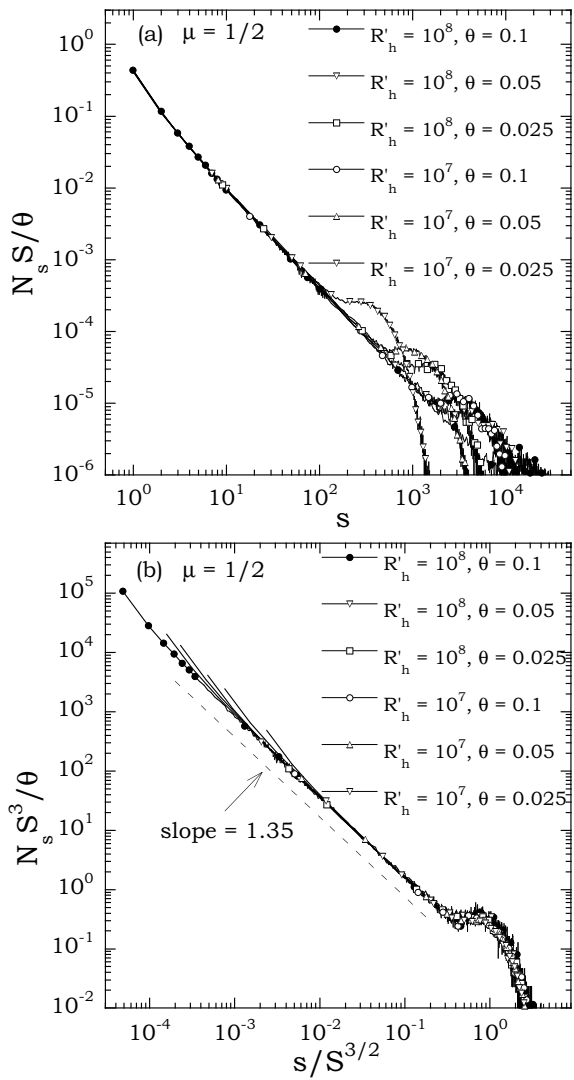}
\caption{(a) Scaled ISD for   $\mu = 1/2$  obtained using steady-state scaling form Eq.~\ref{steadystate2}.  (b) Scaled ISD obtained using generalized scaling form Eq.~\ref{NS2Q} with $\tau = 4/3$ and $\zeta = 3/2$.
}
\label{Fig:isd05}\end{figure}

Fig.~\ref{Fig:isd05}(a) shows the ISD scaled using this form.  As can be seen there is  reasonably good scaling  for $s < S_c$, although   the tail of the distribution does not scale.  However,  the measured value of the exponent $\tau$ ($\tau \simeq 4/3$) is significantly higher than the predicted value.    In addition,
the measured value of $\zeta$ ($\zeta \simeq 3/2$) is also significantly higher than the predicted value.
Fig.~\ref{Fig:isd05}(b) shows the  corresponding scaling results obtained using the generalized scaling form   Eq.~\ref{NS2Q} and assuming $\zeta = 3/2$ and $\tau= 4/3$.  We note that this implies that,
\begin{equation}
N_s(\theta) \sim S^{-3} \theta ~ f(s/S^{3/2})
\label{scale1}
\end{equation}
As can be seen, in this case both the power-law region for small $s/S$ as well as the `bump' for large $s/S$ scale well using this form.
We note however, that   for the smallest clusters (e.g. monomers and dimers)  there is poor scaling due to deviations from power-law behavior for small $s$.

\subsection {Correlated attachment-detachment ($\mu = 1$)}

\begin{figure}[t]
\includegraphics [width=7.5cm]{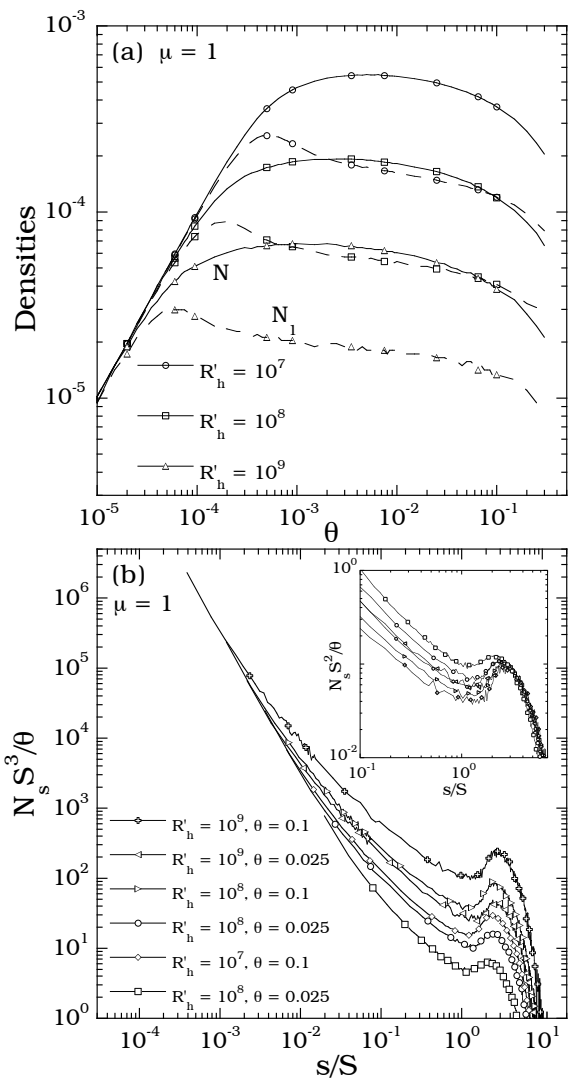}
\caption{(a) Island and monomer densities $N$ and $N_1$  as a function of coverage $\theta$ for ${R'_h}=10^7 - 10^9$ and $\mu = 1$. (b)Scaled ISD for $\mu = 1$   using generalized scaling form (\ref{NS2Q}) with $\zeta = 1$ and $\tau = 2$.   Results correspond to coverages $\theta=0.025,~0.05~,0.1$, ${R'_h}=10^7 - 10^9$ and $\mu=1$.  Inset shows corresponding scaling results obtained using the standard scaling form (\ref{isdscal}). }
\label{Fig:dens1}
\end{figure}

We now consider the case $\mu  = 1$ which corresponds to cluster diffusion via correlated attachment-detachment.  We note that this is the critical value for   power-law behavior of the ISD (which is expected to occur for $0 \le \mu < 1$) and as a result Krapivsky et al\cite{Krapivsky1,Krapivsky2} have predicted ``nested" logarithmic behavior for the island-density.   Since the simulations are not as computationally demanding as for  $\mu = 1/2$, in  this case we have carried out simulations up to $\theta = 0.3$.
Fig.~\ref{Fig:dens1}(a) shows our results for the     total island density $N$ and monomer density $N_1$ as a function of coverage for $R'_h =10^7 - 10^9$.    As can be seen,  while there is a plateau in the island-density which appears to broaden and flatten somewhat with increasing ${R'_h}$, the plateau is not as flat as for the case $\mu = 1/2$, thus indicating deviations from steady-state behavior.
As for the case $\mu=1/2$, a plot   of the scaled densities   $N {R'_h}^{\chi}$ ($N_1{R'_h}^{\gamma}$)  as a function of scaled coverage  
$\theta {R'_h}^{\chi}$  ($\theta {R'_h}^{\gamma}$) 
shows relatively good scaling up to the coverage corresponding to the peak island-density, although the value of $\chi$ ($\chi \simeq 0.45$) is slightly lower than that obtained for $\mu = 1/2$.

We now consider the island-size distribution.  As shown in Fig.~\ref{Fig:dens1}(b), in this case the ISD does not exhibit a well-defined power-law behavior.  In particular, on a log-log plot the ISD is curved  with a slope $\tau \simeq 2$ for small $s$ and a smaller effective slope ($\tau \simeq 1$)  for large $s$.
Similarly, while $\zeta \simeq 1$ its effective value ranges from $1.03$ to $1.1$ depending on the value of $R'_h$  and   coverage.  As a result, neither the standard scaling form Eq.~\ref{isdscal} nor the generalized scaling form Eq.~\ref{NS2Q} can be used to scale the entire island-size distribution.  However, using the generalized  scaling form (\ref{NS2Q})  with $\zeta \simeq 1$ and $\tau = 2$, we find  good scaling for small $s/S$ (see Fig.~\ref{Fig:dens1}(b)), although the ISD does not scale for large $s/S$. On the other hand, if we use  the standard scaling form (\ref{isdscal}) (which corresponds to the generalized scaling form with  $\zeta = 1$ and $\tau = 1$, see inset of Fig.~\ref{Fig:dens1}(b)) then the ISD scales for $s > S_c$ but not for small $s$.
We note that this lack of scaling is perhaps not surprising since for $\mu \ge 1$ there are  two characteristic size-scales $S$ and $S_c$,  but  no well-defined power-law behavior.

\subsection {Periphery diffusion ($\mu = 3/2$)}

\begin{figure}[t]
\includegraphics [width=7.5cm]{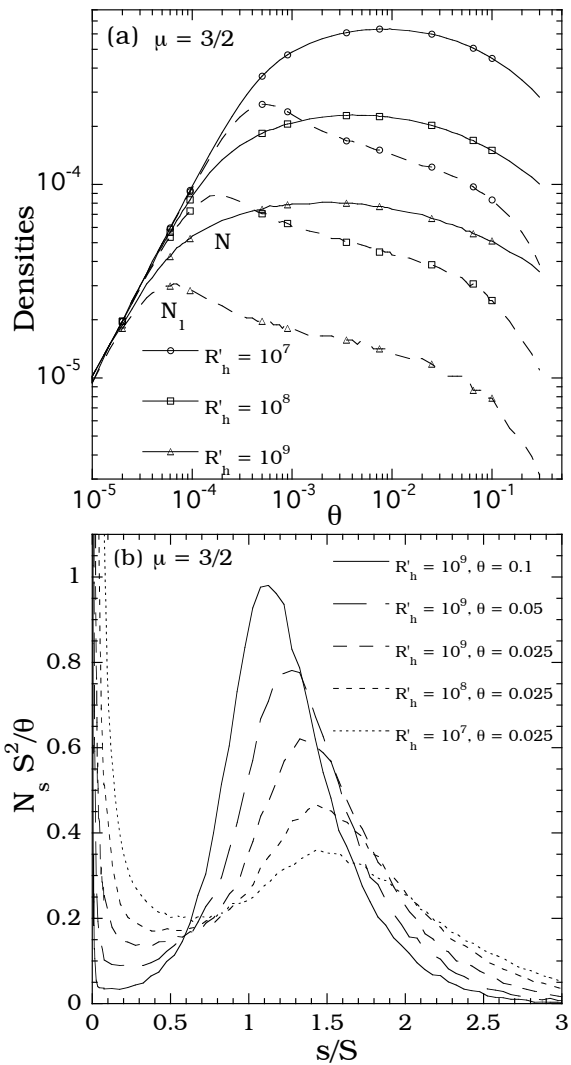}
\caption{(a) Island and monomer densities $N$ and $N_1$  as a function of coverage $\theta$ for ${R'_h}=10^7 - 10^9$ and $\mu = 3/2$. (b) Scaled ISD for $\mu = 3/2$  using standard scaling form (\ref{NS2Q}) with $\zeta = 1$ and $\tau = 2$(\ref{isdscal}).}
\label{Fig:dens32}
\end{figure}

We now consider the case $\mu  = 3/2$ which corresponds to cluster diffusion via edge-diffusion.
Fig.~\ref{Fig:dens32}(a) shows our results for the   total island density $N$ and monomer density $N_1$ as a function of coverage for $R'_h = 10^7 - 10^9$.    As can be seen,  while there is a plateau in the island-density which appears to broaden   with increasing ${R'_h}$, it  is not  as flat as for the case $\mu = 1$, thus indicating deviations from steady-state behavior.
As for the case $\mu=1$, a plot of the scaled densities   $N {R'_h}^{\chi}$ ($N_1{R'_h}^{\gamma}$)  as a function of scaled coverage  
$\theta {R'_h}^{\chi}$  ($\theta {R'_h}^{\gamma}$) 
shows relatively good scaling up to the coverage corresponding to the peak island-density.  

We note that for $\mu > 1$,  Krapivsky et al\cite{Krapivsky1,Krapivsky2} have predicted that for point-islands there is a continuous logarithmic increase in the total island density of the form,
\begin{equation}
N  \simeq   R^{-1/2} \left[\frac  {\sin(\pi/\mu)}{\pi} \ln(\theta R^{1/2})\right]^{\mu/2}
\label{sinequ}
\end{equation}
However, we find that for $\mu = 3/2$ and higher (not shown)   scaling plots using this form (e.g. $N R^{1/2}$ as a function of $[\ln(\theta R^{1/2})]^{\mu/2}$)   provide very poor scaling.  In particular, since $\chi \simeq 0.45$, the scaled peak island-density increases with $R$ while the peak position also shifts significantly to smaller values.

We now consider the scaled ISD for $\mu = 3/2$.  Again in this case, it is not possible to scale the entire ISD using the average island-size $S$ since there are two characteristic size-scales but no well-defined power-law behavior.
In particular,  if we use the generalized scaling form Eq.~\ref{NS2Q} with $\tau = 2$ 
and $\zeta = 1$, then  reasonable scaling is only obtained for the small-$s$ ``tail" corresponding to $s/S < 0.1$ (not shown).  
In addition, as shown in Fig.~\ref{Fig:dens32}(b),   using the standard scaling form Eq.~\ref{isdscal} neither the tail nor the peak scale.
 We note that the height and width of the ``power-law" portion of the ISD    decreases with increasing ${R'_h}$ and coverage, while the peak near $s/S = 1$ becomes higher and sharper.   As a result, the power-law portion of the ISD
is significantly less important than for smaller values  of $\mu$.  In particular,
for ${R'_h} = 10^9$ and $\theta = 0.1$,  it corresponds to only approximately $10\%$ of the area under the curve.

\begin{figure}[t]
\includegraphics [width=7.5cm]{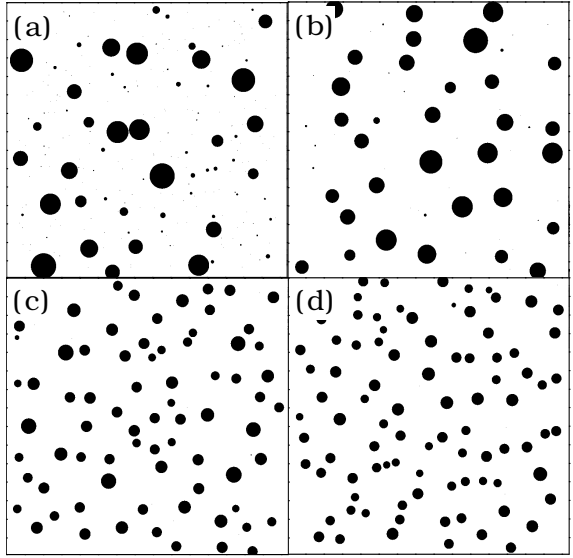}
\caption{Pictures (size $M \times M$) of the submonolayer morphology at  coverage $\theta = 0.1$ and ${R'_h} = 10^9$ for (a) $\mu = 1/2$ (M =  4096) (b) $\mu = 1$ (M =  709) (c) $\mu = 3/2$ (M =  624)   (d) $\mu =  2$ (M =  485).}
\label{Fig:picture}\end{figure}

Fig.~\ref{Fig:picture} shows    pictures of the submonolayer morphology  for $R'_h = 10^9$ and   $\theta = 0.1$  for $\mu = 1/2, 1, 3/2$, and $2$.   We note that the size-scale $M$ of each picture decreases with increasing $\mu$ so that approximately the same number of islands is visible.  As can be seen, in qualitative  agreement with our results, there is a very broad distribution of island-sizes for $\mu = 1/2$ while the distribution becomes narrower with increasing $\mu$.

\subsection {Scaling of ISD and densities for $\mu \ge 2$}

\begin{figure}[t]
\includegraphics [width=7.5cm]{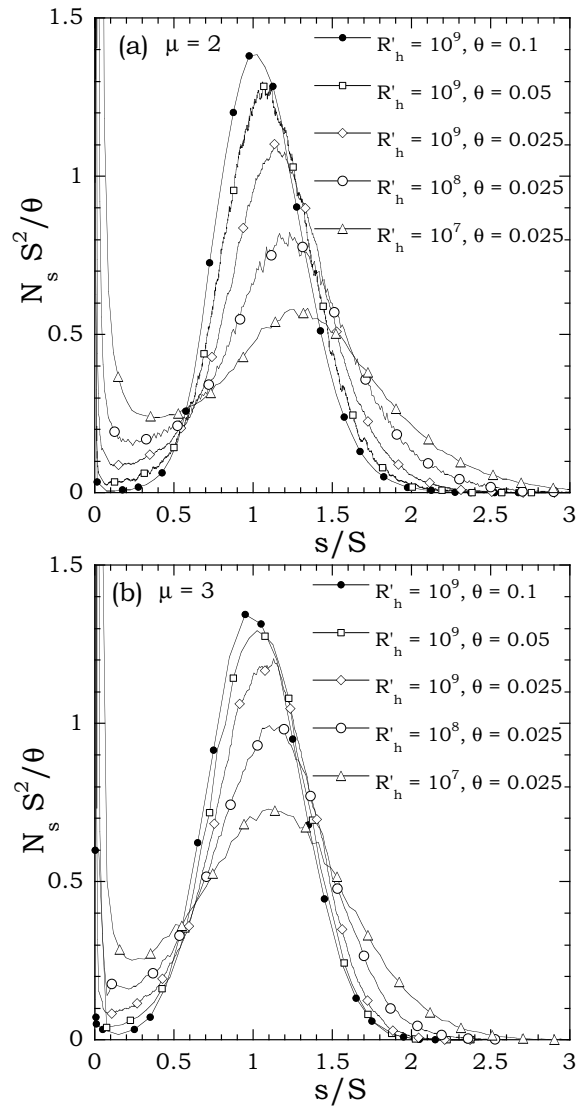}
\caption{Scaled ISD for ${R'_h}=10^7 - 10^9$ and coverage $\theta=0.025-0.1$ for  (a)  $\mu=2$  and (b) $\mu = 3$.
}
\label{Fig:isd2}\end{figure}

In order to obtain a better understanding of  the dependence of the island density and ISD on the mobility exponent $\mu$, we have also carried out additional simulations for larger values of $\mu$ ($\mu = 2, 3$ and $6$)    as well as in the limit  $\mu = \infty$ in which only monomers can diffuse.
Fig.~\ref{Fig:isd2}(a) shows the corresponding results for the scaled ISD for $\mu = 2$  using the standard scaling form Eq.~\ref{isdscal}  for different values of the coverage $\theta$ and  $R'_h$.   As for the case $\mu = 3/2$ the ISD does not scale, although the
   ``power-law" portion for small $s/S$ is significantly reduced.  Instead the peak of the scaled ISD  increases with increasing coverage and $R'_h$.     We also note that for $R'_h = 10^9$ and $\theta = 0.1$,  the peak height is significantly higher than for $\mu=3/2$ while the peak position  is closer to $s/S=1$.
   Similar results for the scaled ISD for $\mu = 3$ are shown in Fig.~\ref{Fig:isd2}(b), although in this case it tends to sharpen more rapidly with increasing  $R'_h$ and coverage.
These results also suggest that,  while the scaled ISD {\it may} approach a well-defined form (independent of coverage and $R'_h$)  in the asymptotic limit of  large $R'_h$, the corresponding scaling function   depends on $\mu$.  Such a $\mu$-dependence is  consistent with the dependence of the exponent $\chi$ and $\chi'$ on $\mu$ (see Fig.~\ref{Fig:allchi}).

\begin{figure}[t]
\includegraphics [width=7.5cm]{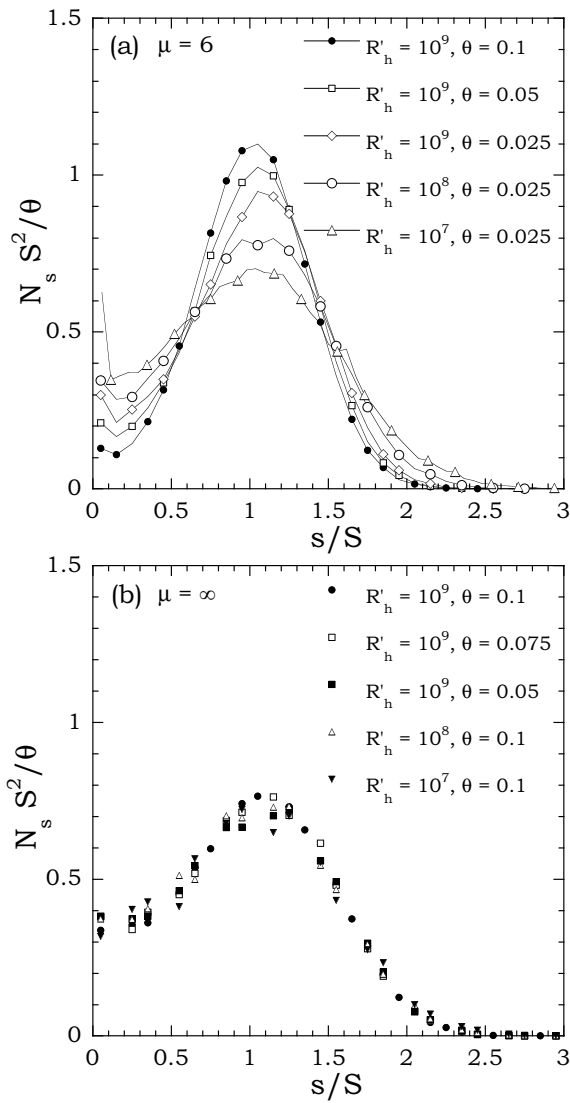}
\caption{Scaled ISD for ${R'_h}=10^7 - 10^9$ and coverage $\theta=0.025-0.1$ for  (a)  $\mu=6$ and  (b) $\mu = \infty$.}
\label{Fig:isd6}\end{figure}

Fig.~\ref{Fig:isd6} shows  our results for the scaled ISD for $\mu = 6$ as well as in  the limit  $\mu =  \infty$ in which only monomers can diffuse.  Somewhat surprisingly, we find that for $\mu = 6$
the scaled ISD   is significantly broader than for $\mu =2$ and $\mu = 3$, although it is still more sharply-peaked than   for $\mu = \infty$.  These results suggest that, at least for (finite) fixed $R'_h$,  the  peak-height depends non-monotonically on $\mu$, e.g. it increases from $\mu = 3/2$ to $\mu = 3$ but  then decreases for higher $\mu$.  This is  also consistent with our results for $\mu = \infty$ (see Fig.~\ref{Fig:isd6}(b)) for which   good scaling is observed  but with a peak height which is lower than for $\mu = 6$.

Fig.~\ref{Fig:allchi}(a) shows a summary of our results for the monomer density $N_1$ and total island density $N$ as a function of coverage for $\mu = 1/2, 1, 3/2, 2, 6$, and $\infty$ for the case $R'_h = 10^9$.  As can be seen, up to the coverage $\theta_{1,m}$ corresponding to the peak monomer density both the island and monomer density are essentially independent of $\mu$.  Fig.~\ref{Fig:allchi}(a) also shows clearly that both the island-density and the coverage $\theta_m$ corresponding to the peak island-density increase with increasing $\mu$, while the   monomer density  decreases with increasing $\mu$.

Fig.~\ref{Fig:allchi}(b)
 shows a summary of our results for the dependence of the exponents $\chi$, $\chi'$, and $\gamma$ on $\mu$.
As can be seen, the exponent $\chi$ depends continuously on $\mu$, decreasing from a value close to $1/2$ for small $\mu$ ($\mu = 1/2$) and approaching a value close to $1/3$ for large $\mu$.  We note that these results are similar to previous results obtained for fractal islands with $d_f = 1.5$ by Mulheran and Robbie.\cite{Mulheran}
Similarly, we find that  the exponent $\chi'$ describing the dependence of the monomer density  at fixed coverage  on $R'_h$ also shows a continuous variation with increasing $\mu$, starting at a value close to $1/2$  for $\mu = 1/2$  and increasing to a  value close to $2/3$ for large $\mu$. In contrast, the exponent $\gamma$ describing the flux-dependence of the peak monomer density is  close to $1/2$ for all $\mu$. 

\section{Discussion}

Motivated in part by recent experiments on colloidal nanoparticle island nucleation and growth during droplet evaporation,\cite{Bigioni}
we have carried out   simulations of a simplified model of irreversible  growth of compact islands in the presence of monomer  deposition  and a power-law dependence ($D_s \sim s^{-\mu}$) of the island mobility $D_s$ on island-size $s$.  In particular, we have considered the cases $\mu = 1/2$ (corresponding to cluster-diffusion via Brownian motion),  $\mu = 1$  (corresponding to cluster-diffusion via correlated evaporation-condensation),  and $\mu = 3/2$ (corresponding to cluster-diffusion via periphery diffusion).  For comparison, we have also carried out simulations for higher values of $\mu$ including  $\mu = 2, 3$ and $6$ as well as $\mu = \infty$.

\begin{figure}[t]
\includegraphics [width=7.5cm]{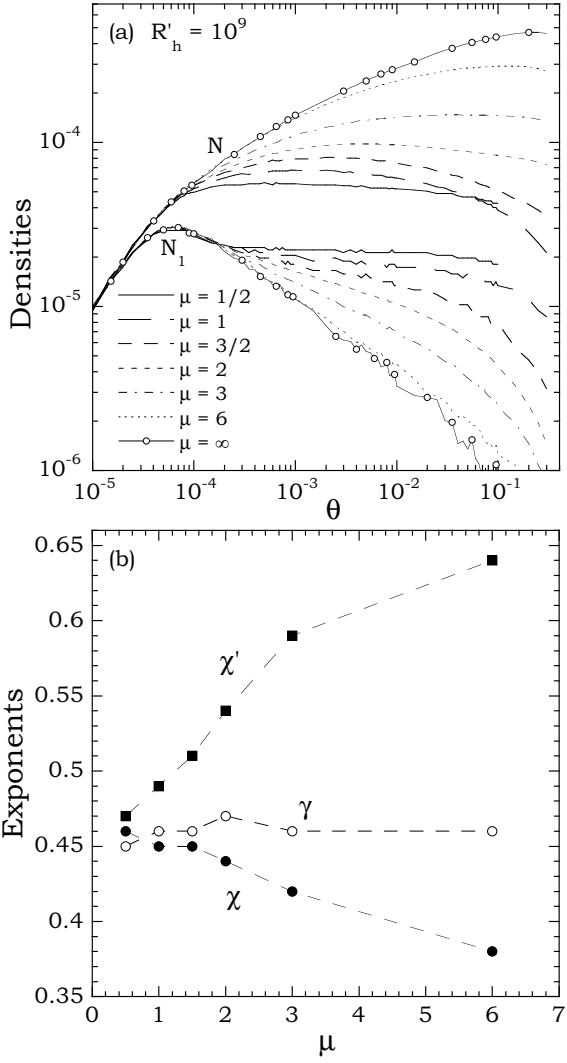}
\caption{(a) Island density $N$ and monomer density $N_1$ as function of coverage for $R'_h = 10^9$ and $\mu = 1/2, 2, 3, 6$, and $\infty$.  (b) Dependence of   exponents $\chi$, $\chi'$, and $\gamma$ on the parameter $\mu$.}
\label{Fig:allchi}\end{figure}

In agreement  with the  predictions of Ref.~\onlinecite{Krapivsky1} and   Ref.~\onlinecite{Krapivsky2} for point-islands, we find that for small values of $\mu$
 the value of  the exponent $\chi$ characterizing the dependence of the peak-island density on $R'_h$ is close to but slightly lower than $1/2$.   However,   we also find
that $\chi$ decreases continuously with   increasing $\mu$,  approaching the value $1/3$ for large $\mu$.
As already noted, these results are in good agreement with previous results obtained for fractal islands.\cite{Mulheran}
Similarly,   the exponent $\chi'$ characterizing the dependence of the peak monomer density on $R'_h$ is also close to $1/2 $ for small $\mu$,  but increases with increasing $\mu$, approaching the value $2/3$ in the limit $\mu \rightarrow \infty$.  In contrast,   the exponent $\omega$ describing the dependence of the coverage $\theta_m$ (corresponding to the peak-island density)  on $R'_h$ is significantly smaller than $1/2$ for small $\mu$ and also decreases with $\mu$,  approaching zero in the limit of infinite $\mu$.  This is consistent with the fact that when only monomers are mobile ($\mu = \infty$)  the peak island-density occurs at a coverage which is independent of $R'_h$ in the asymptotic  limit of large $R'_h$.
For comparison, we note that while the monomer density $N_1(\theta)$ depends on $R'_h$ it only depends on $\mu$ for coverages {\it beyond}  the peak monomer density (see Fig.~\ref{Fig:allchi}(a)).
As a result, the exponents $\gamma$ and $\omega'$ corresponding to the dependence of the peak monomer density (and  corresponding coverage $\theta_{1,m}$)   on $R'_h$
are  close to $1/2$ for all $\mu$.

The similarity of our results for $\chi$ and $\omega$ to previous  results\cite{Mulheran} for fractal islands suggests that these exponents (along with the exponent $\chi'$) depend primarily on the cluster-mobility exponent $\mu$ and substrate-dimension $d$ but not on the shape or fractal dimension of the islands. 
We note that such a result is not entirely surprising, since for the   case in which only monomers can diffuse ($\mu = \infty$) it has been found that the exponent $\chi$ depends only weakly on the island fractal dimension.\cite{Evansreview}
In addition, we have found that the scaled island and monomer densities ($N R^\chi$ and $N_1 R^\chi$) lead  to reasonably good scaling as a function of $\theta R^\chi$, up to and somewhat beyond the peak island-density.   We note that this scaling form is somewhat different from that used in Ref.~\onlinecite{Mulheran} in which the coverage is scaled by $\theta R^\omega$ so that only the peak scales.

 In addition to the scaling of the island and monomer densities, we have also studied the dependence of the island-size distribution (ISD) on the cluster-mobility exponent $\mu$.    In agreement with  the  prediction\cite{Krapivsky1, Krapivsky2,Cueille, Camacho}
  that for point-islands
  well-defined power-law behavior should be observed for    $\mu < 1$, for the case $\mu = 1/2$ we find a broad distribution of island-sizes with a well-defined power-law.  However, in contrast to the point-island prediction that $\tau = (3-\mu)/2$ (which implies $\tau = 5/4$ for $\mu = 1/2$) the value of $\tau$ obtained in our simulations ($\tau \simeq 4/3$) is somewhat larger.  Similarly, the value of the exponent ($\zeta \simeq 3/2$)  describing the dependence of the crossover island-size $S_c$ on $S$ for $\mu = 1/2$  is also significantly larger than the point-island prediction $\zeta = 2/(\mu +1) = 4/3$.
One possible explanation for this is that  for compact islands the coalescence rate decreases more slowly with increasing island-size  than for point-islands due to the increase in ``aggregation cross-section" with increasing island-radius. However, another possible explanation is the existence of correlations that are not included in the mean-field Smoluchowski equations.  In particular, we note that in previous work for the case of irreversible growth in the absence of cluster diffusion ($\mu = \infty$), it has been shown\cite{Bartelt} that there exist strong correlations between the size of an island and the surrounding capture-zone.  

\begin{figure}[t]
\includegraphics [width=7.5cm]{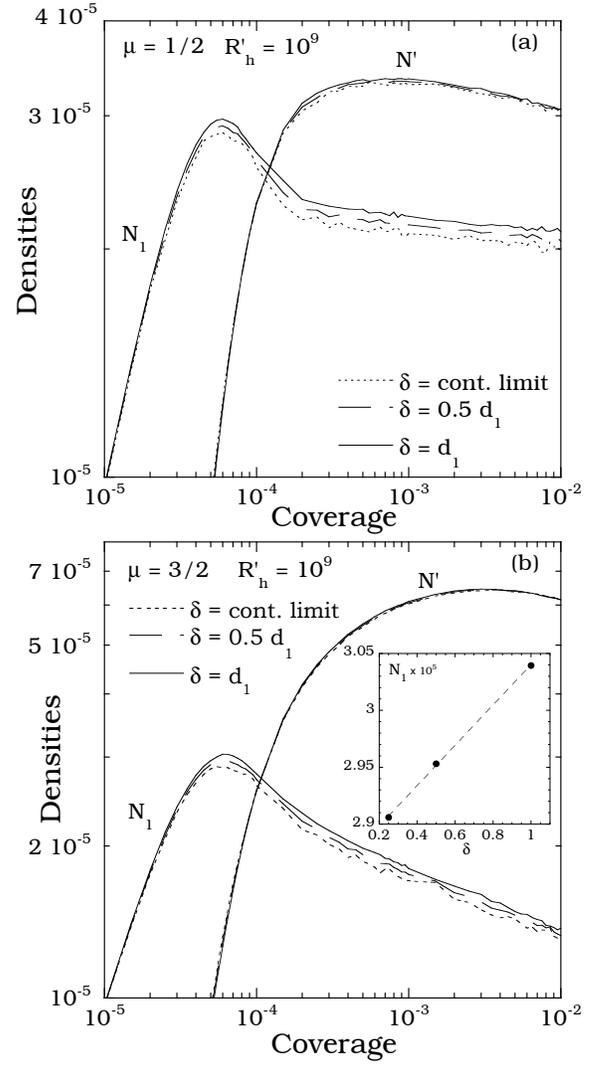}
\caption{Island density $N'$ (not including monomers) and monomer density $N_1$ for $\delta=d_1$, $\delta=0.5 d_1$, and continuum limit corresponding to $\delta = 0$. 
Inset shows dependence of peak monomer density $N_1$  on  $\delta$.}
\label{Fig:delta}\end{figure}

 We note that  in contrast to previously studied growth models with  only limited cluster-diffusion,\cite{Evansreview,BE94,Amar94,iprl}
 in which there is a single well-defined peak in the ISD corresponding  to the average island-size $S$,
 in the presence of significant cluster mobility
 there are typically  two different size-scales  $S$ and $S_c$.  As a result,  in general it  is not  possible to scale the ISD using just the average island-size $S$.  However, for the case $\mu < 1$ (corresponding to well-defined power-law behavior up to a critical island-size $S_c$)   our results confirm that for compact  islands the ISD exhibits steady-state behavior.  As a result,   the power-law region corresponding to $s < S_c$  can be scaled using Eq.~\ref{steadystate2}, although the large-$s$ ``tail" does not scale.
  Accordingly, we have proposed a generalized   scaling form for the ISD,  $N_s(\theta)  = \theta/S^{1+\tau \zeta}   f(s/S^\zeta)$
  for the case $\mu < 1$.
Using this form, we have obtained excellent scaling
for  the case $\mu = 1/2$.

 In contrast for $\mu = 1$,
 there are still two competing size-scales $S$ and $S_c$, but  there is no well-defined power-law behavior.  As a result, no single scaling form can be used to scale the entire ISD.  However, we find that
  the value of the exponent $\zeta$ ($\zeta \simeq 1$) is close to that obtained using the point-island expression $\zeta = 2/(\mu + 1)$.
 In addition,  for small $s/S$ the ISD satisfies   $N_s(\theta) \sim s^{-\tau_{eff}}$ where $\tau_{eff} \simeq 2$.   As a result, we find that   the small $s/S$  ``tail" of the ISD can be scaled using the generalized scaling form Eq.~\ref{NS2Q} with $\tau = 2$ and $\zeta = 1$, while  the standard scaling form Eq.~\ref{isdscal} leads to reasonably good scaling of the ISD for   $s > S_c$.

However, for $\mu > 1$   there is no effective power-law behavior and as a result, neither the general scaling form Eq.~\ref{NS2Q} nor the standard scaling form Eq.~\ref{isdscal}   lead to good scaling of the ISD for finite $R'_h$.    Instead we find that, using the standard scaling form Eq.~\ref{isdscal}   the fraction of islands corresponding to small $s/S$ decreases with increasing $R'_h$, and coverage, while the peak of the scaled ISD  increases in height and becomes sharper. As a result,
the peak position shifts to the left  with increasing $R'_h$ and coverage and appears to approach $1$ for large $R'_h$.
Interestingly, this implies, as shown in Figs.~\ref{Fig:dens32}(b), \ref{Fig:isd2}, and \ref{Fig:isd6},  that for $\mu > 1$ the peak of the scaled ISD is even higher than for the case of irreversible growth without cluster diffusion ($\mu = \infty$). However, our results also suggest that, at least for fixed coverage and  finite (fixed) $R'_h$, the   peak-height of the scaled ISD exhibits a non-monotonic dependence on $\mu$, since it increases from $\mu = 3/2$ to $\mu = 3$ but is smaller for $\mu = 6$.

It is also interesting to compare our results for $\mu > 1$ with those obtained by Kuipers and Palmer\cite{Kuipers} who studied the scaled ISD for the case of fractal islands, assuming an exponential dependence of the cluster mobility, e.g. $D_s \sim D_1 \xi^s$ where $\xi < 1$.  Because of the rapid decay of the mobility with increasing cluster-size   assumed in their model, the resulting scaled island-size distributions (using the standard scaling form Eq.~\ref{isdscal}) were much closer to those obtained for the case of irreversible growth with no cluster mobility (e.g. $\mu = \infty$)  than   the results presented here.  However, for values of $\xi$ which were not too small, they also found some evidence of a small island-size ``tail", although it was much weaker than  found here.

It is also interesting to
consider the applicability of the model studied here to  recent experiments by Bigioni et al\cite{Bigioni} for the case of colloidal nanoparticle cluster formation during drop-drying.
We note that in this case, one expects that clusters will diffuse on the droplet surface via Brownian motion which implies that $\mu = 1/2$.
However, one also expects that, due to the relatively weak Van der Waals attraction between nanoparticles,  in this case cluster formation may be reversible.
Accordingly, it would be interesting to carry out additional simulations
for the case of reversible growth corresponding to a critical island-size $i \ge 2$.

Finally, we consider the continuum limit of our simulations.
As already mentioned, while our simulations are off-lattice,  in all of the results presented so far  we have assumed a hopping length $\delta$ equal to the monomer diameter $d_1$.  
We note that   this makes our simulations similar to  previous simulations\cite{Evansreview, BE94, Amar94, iprl, Jensen, Mulheran, Krapivsky2}  with and without cluster mobility in which a lattice was assumed.  However, it is also interesting to consider the continuum limit $\delta \rightarrow 0$.
In order to do so, we have carried out additional simulations with smaller values of $\delta$ ($\delta = d_1/2$ and $d_1/4$). In general, we find that both the monomer density $N_1$, as well as the density $N'$ of all clusters  not including monomers    exhibit a  weak but linear dependence on the hopping length  $\delta$  (see inset of Fig.~\ref{Fig:delta}(b)) e.g.,
\begin{equation}
X(\delta) = X(0) [1+ \alpha(\mu)~ (\delta/d_1)]
\end{equation}
(where $X$ corresponds either to  the monomer or island density and $X(0)$ corresponds to the continuum limit).
Accordingly, by performing a linear extrapolation 
we may obtain the corresponding densities in the continuum limit. As shown in Fig.~\ref{Fig:delta}, for $\mu = 1/2 $ and $\mu = 3/2$ the island-density $N'$   depends relatively weakly on the hopping length, and as a result there is very little difference between our results for $\delta = d_1$ and the continuum limit.
In contrast,   the monomer density exhibits a somewhat stronger dependence on the hopping length $\delta_1$.
However, in general we find $\alpha (\mu) < 0.1$ while
the value of $\alpha(\mu)$ decreases with increasing $\mu$.  In particular, in the limit  $\mu = \infty$ in which only monomers can diffuse, we find $\alpha(\infty) = 0.01$ ($0.07$) for the island and monomer density respectively.
These results indicate that in the continuum limit the island and monomer densities are only slightly lower than in our simulations.  Accordingly, we expect that in the continuum limit the scaling behavior will not be significantly different from the results presented here.

\bigskip

\begin{acknowledgments}
This work was supported by Air Force Research Laboratory, Space
Vehicles Directorate (Contract No. FA9453-08-C-0172) as well as by NSF grant CHE-1012896.  
We would also like to acknowledge  a grant of computer time from the Ohio Supercomputer Center.
\end{acknowledgments}


\begin{thebibliography}   {99}
\bibitem{Kuipers-sim}  Kuipers and Palmer\cite{Kuipers} have considered the scaling of the ISD in a similar model. However, in their model the cluster diffusivity was assumed to decay   exponentially   with cluster-size.
\end{thebibliography}
\end{document}